\documentclass[prl,twocolumn,longbibliography,nofootinbib, preprintnumbers,notitlepage,fleqn,showpacs,superscriptaddress]{revtex4-1}

\usepackage{color,soul}

\usepackage{contour}
\usepackage{ulem}

\contourlength{0.8pt}

\newcommand{\myuline}[1]{%
  \uline{\phantom{#1}}%
  \llap{\contour{white}{#1}}%
}
\renewcommand{\emph}{\textit}

\usepackage[bookmarks = true, citecolor = blue, colorlinks = true, linkcolor = magenta, urlcolor = blue]{hyperref}

\usepackage[T1]{fontenc}			
\usepackage[sc,osf]{mathpazo}   	

\usepackage{amsmath}  			
\usepackage{amsfonts}  			
\usepackage{graphicx}   			
\usepackage{mathrsfs, amsthm, amssymb}
\usepackage{bbm, bm}

\usepackage{nicefrac}    			

\usepackage[braket, qm]{qcircuit}	

\usepackage{tikz}
\usetikzlibrary{calc,decorations.pathmorphing,shapes}

\newcounter{sarrow}
\newcommand\xrsquigarrow[1]{%
\stepcounter{sarrow}%
\mathrel{\begin{tikzpicture}[baseline= {( $ (current bounding box.south) + (0,-0.5ex) $ )}]
\node[inner sep=.5ex] (\thesarrow) {$\scriptstyle #1$};
\path[draw,<-,decorate,
  decoration={zigzag,amplitude=0.7pt,segment length=1.2mm,pre=lineto,pre length=4pt}] 
    (\thesarrow.south east) -- (\thesarrow.south west);
\end{tikzpicture}}%
}

\makeatletter
\newcommand{\dbloverline}[1]{\overline{\dbl@overline{#1}}}
\newcommand{\dbl@overline}[1]{\mathpalette\dbl@@overline{#1}}
\newcommand{\dbl@@overline}[2]{%
  \begingroup
  \sbox\z@{$\m@th#1\overline{#2}$}%
  \ht\z@=\dimexpr\ht\z@-2\dbl@adjust{#1}\relax
  \box\z@
  \ifx#1\scriptstyle\kern-\scriptspace\else
  \ifx#1\scriptscriptstyle\kern-\scriptspace\fi\fi
  \endgroup
}
\newcommand{\dbl@adjust}[1]{%
  \fontdimen8
  \ifx#1\displaystyle\textfont\else
  \ifx#1\textstyle\textfont\else
  \ifx#1\scriptstyle\scriptfont\else
  \scriptscriptfont\fi\fi\fi 3
}
\makeatother

\begin{document}


\title[]{Efficient linear optical generation of a multipartite $\mathbf{W}$ state}



\author{Pawel \surname{Blasiak}}
\email{pawel.blasiak@ifj.edu.pl}
\affiliation{Institute of Nuclear Physics Polish Academy of Sciences, PL-31342 Krak\'ow, Poland}
\affiliation{City, University of London, London EC1V 0HB, United Kingdom}
\author{Ewa \surname{Borsuk}}
\email{ewa.borsuk@ifj.edu.pl}
\affiliation{Institute of Nuclear Physics Polish Academy of Sciences, PL-31342 Krak\'ow, Poland}
\author{Marcin \surname{Markiewicz}}
\email{marcinm495@gmail.com}
\affiliation{International Centre for Theory of Quantum Technologies, University of Gda\'nsk, PL-80308 Gda\'nsk, Poland}
\author{Yong-Su \surname{Kim}}
\email{yong-su.kim@kist.re.kr}
\affiliation{Center for Quantum Information, Korea Institute of Science and Technology (KIST), Seoul 02792, Republic of Korea}
\affiliation{Division of Nano \& Information Technology, KIST School, Korea University of Science and Technology, Seoul 02792, Republic of Korea}


\begin{abstract}
A novel scheme is presented for generation of a multipartite $W$ state for arbitrary number of qubits. Based on a recent proposal of entanglement without touching, it serves to demonstrate the potential of particle indistinguishability as a useful resource of entanglement for practical applications. The devised scheme is efficient in design, meaning that it is built with linear optics without the need for auxiliary particles nor measurements. Yet, the success probability is shown to be highly competitive compared with the existing proposals (i.e. decreases polynomially with the number of qubits) and remains insensitive to particle statistics (i.e. has the same efficiency for bosons and fermions).
\end{abstract}


\maketitle

\textit{Introduction.}---Entanglement is a key quantum information resource which provides a basis for all modern developments in quantum technologies and remains a central theme in quantum foundations research~\cite{Be93,NiCh00,HoHoHoHo09,BrCaPiScWe14,GiRiTiZb02,LaJeLaNaMoOB10,AsWa12,PaChLuWeZeZu12,WaScLaTh20}. Flexibility of entanglement generation is thus crucially important for practical implementations which require manipulation of an increasing number of qubits prepared in a desired state. Needless to say that the challenges lay on both experimental as well as theoretical side, with various factors deciding about the usefulness of a given proposal. It is always a subtle interplay between the scaling properties for the efficiency of state generation and the resources measured by the complexity of the experimental design.

Here we focus on a prominent example of a multipartite entangled state, the $W$~state, which for $N$ qubits takes the form 
\begin{eqnarray}\label{W-state}
W_{\scriptscriptstyle N}=\tfrac{1}{\sqrt{N}}\,\big(\!\ket{\uparrow\downarrow\downarrow...\downarrow}+
\ket{\downarrow\uparrow\downarrow...\downarrow}+...+\ket{\downarrow...\downarrow\downarrow\uparrow}\!\big),\ 
\end{eqnarray}
where $\{\ket{\uparrow},\ket{\downarrow}\!\}$ is a computational basis. A remarkable property of this state is that the amount of entanglement shared between any of the qubits and all the rest is optimal, in a sense that entanglement is robust to the loss of one or more qubits in the system~\cite{KoBuIm00} (this should be compared with a multipartite $GHZ$ state for which the loss of a single qubit forfeits any entanglement between all the rest). Notably, the $W$ state belongs to a separate entanglement class under SLOCCs~\cite{DuViCi00}. It also shows stronger non-local properties~\cite{SeSeWiKaZu03}. This makes the multipartite $W$ states an interesting quantum resource, with a promise for practical applications in quantum information~\cite{AgPa06,ZhXuPe15,LiMuWe18}. There have been a few implementations of the $W$ state~\cite{EiKiBoKuWe04,MiLiFuKo05,TaWaOzYaKoIm09,TaKiOzYaKoIm10,FaMeLiSiLo19}, but their complexity which grows with the number of qubits presents a significant challenge. For most theoretical proposals~\cite{ZaYaOzSoCa16,WuJi18,BuOzFeKo20} the efficiency drops exponentially with $N$, with the recent exception in 
Ref.~\cite{KiChLiHa20} (see also Ref.~\cite{BeLoCo17} for fermions). 

In this paper, we develop a new scheme for generation of the multipartite $W$ state which can successfully compete with all known proposals, including the quantum erasure proposal in Ref.~\cite{KiChLiHa20}. Our proposition is based on a novel \textit{no-touching} paradigm for entanglement generation in linear optical circuits explicitly discussed in Ref.~\cite{BlMa19} (see also Refs.~\cite{YuSt92,YuSt92a} for early indications of this idea and Refs.~\cite{NeOfChHeMaUm07,KiPrChYaHaLeKa18,JuYaPaChCa19} for some particular realisations). This idea originates from the foundational question about the possibility of entanglement extraction from pure particle indistinguishability without any component from particle interactions. The problem has led to a special class of linear optical designs in which the particles traversing the circuit never touch one another over the entire evolution. Apparently, such experimental schemes provide an efficient platform for entanglement generation, which in the present paper is illustrated on the example of the multipartite $W$ state. We also note that the no-touching designs are insensitive to particle statistics, that is they are equally suitable for bosons and  fermions.

The paper starts with some brief remarks on various factors relevant for the assessment of experimental proposals for state generation. Then we proceed with a detailed description of the no-touching scheme for generation of the multipartite $W$ state for arbitrary number of qubits. As the basis for our construction we use the dual-rail encoding of qubits and then give the corresponding design for polarisation encoding. We also comment on experimental feasibility of the proposed schemes. The paper concludes with a discussion in which the efficiency of our design is compared with other proposals in the literature.

\textit{Remarks on efficiency comparison.}---A standard tool for comparison of experimental designs is the efficiency denoted here by $\textsl{\textsf{Eff}}_{\scriptscriptstyle N}$, i.e. the success probability of obtaining the desired result (in our case it is a given state). The main factor reducing the efficiency comes form post-selection which occurs in many different guises in virtually every experimental design. A typical example are event-ready techniques  for entanglement generation (e.g. entanglement swapping~\cite{ZuZeHoEk93,PaBoWeZe98,MaHeScWaKrNaWi12}) or more generally techniques based on coincidence counts (e.g.~\cite{ZhLiLiPeSuHuHe18,KrHoLaZe17}). Then an evident figure of merit for multiqubit entanglement generation schemes is the scaling of the efficiency $\textsl{\textsf{Eff}}_{\scriptscriptstyle N}$ with the increasing number of qubits $N$. 

However, such an analysis blatantly ignores other details of the experimental design which may easily compromise the utility of a given proposal. A more thorough discussion should take into account the complexity as well as the resources required to implement a given proposal. Here we briefly point out some relevant factors in experimental designs which may have an impact that affects the overall assessment. 

Suppose we consider a scheme that generates a given quantum state of $N$ qubits which, for simplicity, are encoded in particle degrees of freedom (e.g. polarisation, spin or dual-rail encoding). It means that $N$ particles carry the desired state $\psi_{\scriptscriptstyle out}$ at the output. The efficiency $\textsl{\textsf{Eff}}_{\scriptscriptstyle N}$ describes how often this state is produced, as a result of some well defined post-selection procedure. Then, even if the complexity of linear operations in the circuit is not taken into account, the following features of the design still remain relevant:\vspace{-0.2cm}

\begin{flushleft}
\begin{itemize}
\item[\textit{(a)}] How many particles the scheme starts with? What input state $\psi_{\scriptscriptstyle in}$ are they prepared in? \vspace{-0.2cm}
\item[\textit{(b)}] Are intermediate measurements required in the design? Does the scheme involve feedforward? \vspace{-0.2cm}
\end{itemize}
\end{flushleft}

\myuline{\textit{Ad.\,{(a)}}}. Every state generation scheme requires some initial multi-particle state $\psi_{\scriptscriptstyle in}$ to start with, and some of them are easier to prepare than others. In particular, the less particles the better. Therefore every auxiliary particle in the input, on top of what is gained in the output, should count as an extra cost. Furthermore, if our goal is generation of entanglement, then any scheme starting with separable states $\psi_{\scriptscriptstyle in}$, like single particle inputs $|1\rangle^{\otimes N}$, should be considered less demanding compared to those requiring entanglement to start with.

\myuline{\textit{Ad.\,{(b)}}}. Measurements are a different kind of resource which introduces non-linearity into the system. Thus the number of intermediate measurements adds to the overall cost of a given design. Even if the problems with detection inefficiency is ignored, the measurements introduce stochastic element to the procedure. This either results in extra post-selection (cf. event-ready techniques~\cite{ZuZeHoEk93,PaBoWeZe98}) or requires active correction using  feedforward. Needless to say that the latter presents a considerable technical challenge~\cite{MaHeScWaKrNaWi12}.

Clearly, the analysis solely based on the scaling of the efficiency $\textsl{\textsf{Eff}}_{\scriptscriptstyle N}$ with the increasing number of particles $N$ neglects all other relevant factors like those mentioned above. However those other resources required in the design significantly contribute the overall complexity (or cost) deciding about the experimental utility of a given proposal. This makes it difficult to compare designs utilising different resources, since in general it is not clear how to weigh between their costs. There is however one exception: from two designs having a comparable efficiency $\textsl{\textsf{Eff}}_{\scriptscriptstyle N}$, the one with no extra resources seems to be a better choice. This principle will allow to appreciate the performance of the $W$ state generation scheme constructed in the following section.


\begin{figure}
\centering
\includegraphics[width=\columnwidth]{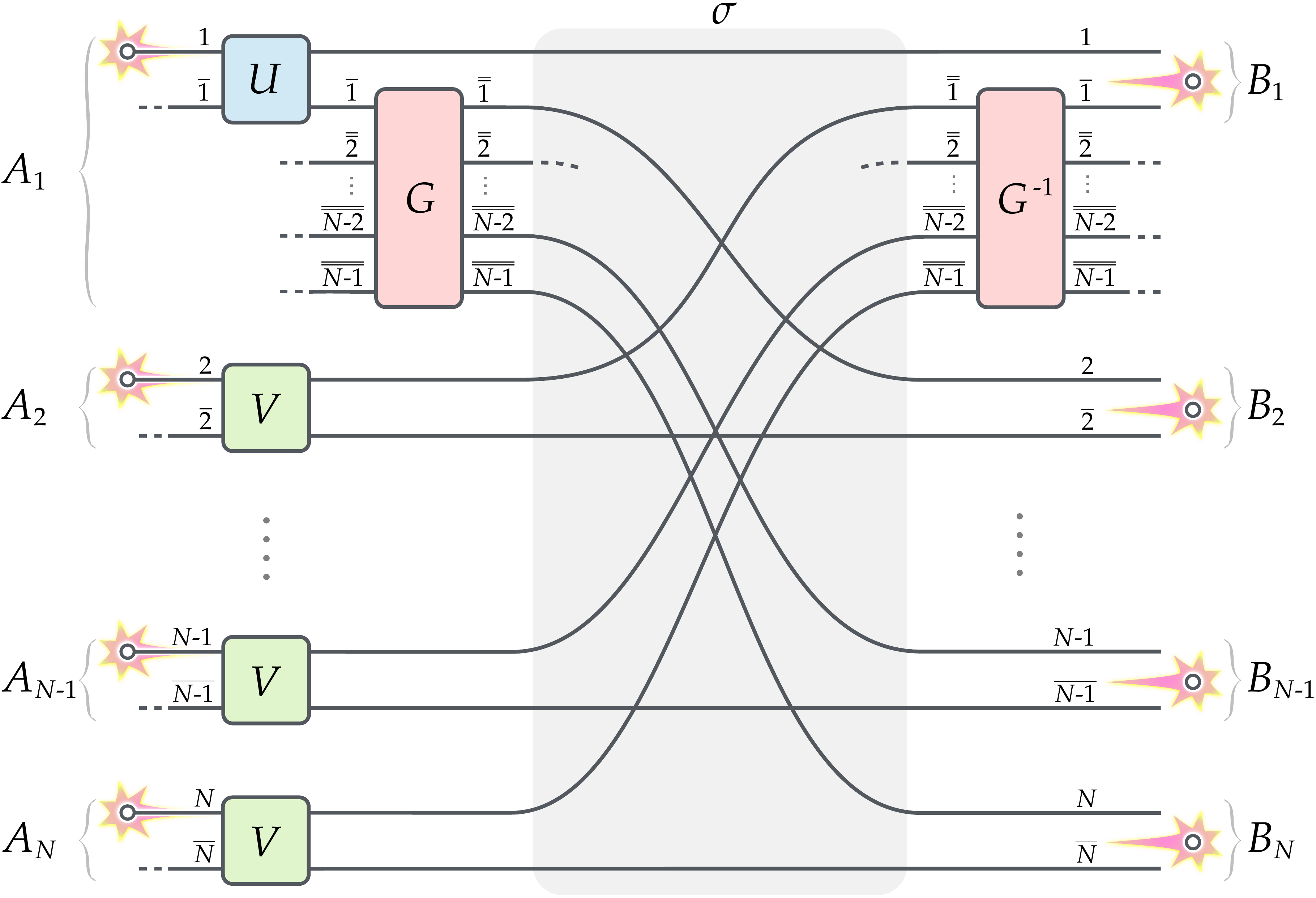}
\caption{\label{Fig_Protocol_Wstate}{\bf\textsf{\mbox{Generation of the $\mathbf{W_{\scriptscriptstyle N}}$ state in the no-touching scheme.}}} N independent particles injected into the circuit undergo a sequence of transformations: local unitaries $U\circ G,V,...\,,V$  in the respective subsystems $A_{\scriptscriptstyle 1}$, ...\,, $A_{\scriptscriptstyle N}$ followed by permutation of the paths $\sigma$ and then local unitary $G^{\scriptscriptstyle{-}1}$ in subsystem $A_{\scriptscriptstyle 1}$. Post-selection on a single particle (coincidence) in each target qubit  $B_{\scriptscriptstyle 1}$, ...\,, $B_{\scriptscriptstyle N}$ generates the $W_{\scriptscriptstyle N}$ for the particular choice of unitaries in Eqs.~(\ref{MatrixUV}) and (\ref{MatrixG}) fulfilling condition Eq.~(\ref{all-equal}).}
\end{figure}

\textit{Preliminaries: No-touching design.}---We follow the discussion of entanglement generation in the no-touching scheme proposed in Ref.~\cite{BlMa19}. Let us consider an optical scheme with $3N-2$ paths (or modes) groupped into $N$ subsystems  $A_{\scriptscriptstyle 1}$, ...\,, $A_{\scriptscriptstyle N}$ with paths labelled as follows:
\begin{eqnarray}\nonumber
\text{System \ $A_{\scriptscriptstyle 1}$}&=&\big\{\,1\,,\overline{1}\,\big\}\cup\big\{\,\dbloverline{2}\,,\dbloverline{3}\,,...\,,\dbloverline{N-1}\,\big\}\,,\\\nonumber
\text{System \ $A_{\scriptscriptstyle 2}$}&=&\big\{\,2\,,\overline{2}\,\big\}\,,\\\nonumber
...\qquad\quad&&\quad...\\\nonumber
\text{System \,$A_{\scriptscriptstyle N}$}\!&=&\big\{\,N\,,\overline{N}\,\big\}\,.
\end{eqnarray}
See Fig.~\ref{Fig_Protocol_Wstate} for illustration. For better clarity we denote with a double bar the modes that will play an auxiliary role in subsystem $A_{\scriptscriptstyle 1}$. Further, we assume that \textit{target} qubits $B_{\scriptscriptstyle 1}$, ...\,, $B_{\scriptscriptstyle N}$ are encoded in pairs of paths in the output (so called \textit{dual-rail} qubits) for which we choose:
\begin{eqnarray}\nonumber
\text{Qubit \ $B_{\scriptscriptstyle k}$}&=&\big\{\,k\,,\overline{k}\,\big\}\qquad\text{for \ $k=1,...\,,N$}\,.
\end{eqnarray}
This means that the computational basis for qubit $B_ {\scriptscriptstyle k}$ is identified with a single particle in the respective path $\ket{\uparrow}_{\scriptscriptstyle k}\equiv a_{\scriptscriptstyle k}^\dag\ket{0}$ and $\ket{\downarrow}_{\scriptscriptstyle k}\equiv a^\dag_{\scriptscriptstyle \overline{k}}\ket{0}$, where $\ket{0}$ is the vacuum. Thus the general state of the qubit is encoded with a single particle in $B_ {\scriptscriptstyle k}$ as $\alpha\,\ket{\uparrow}_{\scriptscriptstyle k}+\beta\ket{\downarrow}_{\scriptscriptstyle k}\equiv(\alpha\,a_{\scriptscriptstyle k}^\dag+\beta\,a_{\scriptscriptstyle \overline{k}}^\dag)\ket{0}$. 

The protocol consists of the following stages (Fig.~\ref{Fig_Protocol_Wstate}):\vspace{-0.15cm}
\begin{itemize}
\item[\textit{(1)}] start with $N$ independent particles with a single particle injected into each subsystem $A_{\scriptscriptstyle 1}$, ...\,, $A_{\scriptscriptstyle N}$,\vspace{-0.15cm}

\item[\textit{(2)}] local unitary on each subsystem $A_{\scriptscriptstyle 1}$, ...\,, $A_{\scriptscriptstyle N}$, \vspace{-0.15cm}

\item[\textit{(3)}] permutation of the paths $\sigma$,\vspace{-0.15cm} 

\item[\textit{(4)}] local unitary in the output on subsystem $A_{\scriptscriptstyle 1}$,\vspace{-0.15cm}

\item[\textit{(5)}] post-selection on a  single particle (i.e. coincidence count) in the output qubits $B_{\scriptscriptstyle 1}$, ...\,, $B_{\scriptscriptstyle N}$.
\end{itemize}

Note that such a design guarantees that the particles do not \textit{touch} one another over the entire evolution as discussed in Ref.~\cite{BlMa19}. For the discussion of its applicability to non-locality tests see Ref.~\cite{BlBoMa20}. One particular property of such schemes is that their efficiency $\textsl{\textsf{Eff}}_{\scriptscriptstyle N}$ is insensitive to particle statistics (i.e. the success probability is equal for bosons and fermions).  In the following, we describe the no-touching protocol generating the $W_{\scriptscriptstyle N}$ state Eq.~(\ref{W-state})
in the target qubits $B_{\scriptscriptstyle 1}$, ...\,, $B_{\scriptscriptstyle N}$. 


\textit{Details of the protocol.}---Let us precise the no-touching protocol given in Fig.~\ref{Fig_Protocol_Wstate} and  define three unitary transformations $U$, $V$ and $G$ in the following form:
\begin{eqnarray}\label{MatrixUV}
U\ =\ \begin{pmatrix}
\alpha&\beta\\\beta&-\alpha
\end{pmatrix}\,,
&\qquad&
V\ =\ \begin{pmatrix}
\delta&\varepsilon\\\varepsilon&-\delta
\end{pmatrix}\,,
\end{eqnarray}
and 
\begin{eqnarray}\label{MatrixG}
G\ =\ \tfrac{1}{\sqrt{N-1}}
\left(\begin{array}{llll}
1&\gamma_{\scriptscriptstyle1,2}&...\ &\gamma_{\scriptscriptstyle 1,N-1}\\
1&\gamma_{\scriptscriptstyle2,2}&...\ &\gamma_{\scriptscriptstyle2,N-1}\\
...&...&...\ &...\vspace{0.1cm}\\
1&\gamma_{\scriptscriptstyle N-1,2}&...\ &\gamma_{\scriptscriptstyle N-1,N-1}
\end{array}
\right)\,.
\end{eqnarray}
For simplicity we will assume that $\alpha$, $\beta$, $\delta$ and $\varepsilon$ are real. Since $U$ and $V$ are unitary, we need to have $\beta^2=1-\alpha^2$ and $\varepsilon^2=1-\delta^2$. Numbers $\gamma_{\scriptscriptstyle k,l}$ are chosen arbitrarily so that $G$ is unitary, and they will not play any role in our argument. Permutation of the modes depicted in Fig.~\ref{Fig_Protocol_Wstate} is defined as follows:
\begin{eqnarray}\label{sigma}
\sigma:\ \ \left\{\ 
\begin{array}{lll}1\rightarrow1\,,&&\\
\dbloverline{k}\rightarrow k+1&\ \ \ &\text{for \ $k=1,...\,,N-1$}\,,\\
k\rightarrow \dbloverline{k-1}&\ \ \ &\text{for \ $k=2,...\,,N$}\,,\\
\overline{k}\rightarrow \overline{k}&\ \ \ &\text{for \ $k=2,...\,,N$}\,.
\end{array}
\right.
\end{eqnarray}
In this way we have specified all components of the protocol. For further convenience, let us write out all unitary transformations that will be used in our analysis:
\begin{eqnarray}\label{U}
a^{\dag}_{\scriptscriptstyle 1} &\!\!\!\xymatrix{\ar[r]^{\atop U} &}\!\!\!&\alpha\,a^{\dag}_{\scriptscriptstyle 1}+\beta\,a^{\dag}_{\scriptscriptstyle \overline{1}}\,,
\\\label{V}
a^{\dag}_{\scriptscriptstyle k} &\!\!\!\xymatrix{\ar[r]^{\atop V} &}\!\!\!&\delta\,a^{\dag}_{\scriptscriptstyle k}+\varepsilon\,a^{\dag}_{\scriptscriptstyle \overline{k}}\qquad\text{for \ $k=2,3,...\,,N$}\,,
\\\label{G}
a^{\dag}_{\scriptscriptstyle \overline{1}} &\!\!\!\xymatrix{\ar[r]^{\atop G} &}\!\!\!&\tfrac{1}{\sqrt{N-1}}\,\big(\,a^{\dag}_{\scriptscriptstyle \dbloverline{1}}+a^{\dag}_{\scriptscriptstyle \dbloverline{2}}+...+a^{\dag}_{\scriptscriptstyle \dbloverline{N-1}}\,\big)\,,
\\\label{G-1}
a^{\dag}_{\scriptscriptstyle \dbloverline{k}} &\!\!\!\xymatrix{\ar[r]^{\atop G^{\scriptscriptstyle\,\text{-}1}} &}\!\!\!&\tfrac{1}{\sqrt{N-1}}\,\big(\,a^{\dag}_{\scriptscriptstyle \overline{1}}+\gamma^*_{\scriptscriptstyle k,2}\,a^{\dag}_{\scriptscriptstyle \dbloverline{2}}+...+\gamma^*_{\scriptscriptstyle k,N-1}\,a^{\dag}_{\scriptscriptstyle \dbloverline{N-1}}\,\big)\ \ \ \ \ \ \\\nonumber
&&\qquad\qquad\qquad\ \text{for \ $k=1,3,...\,,N-1$}\,.
\end{eqnarray}
Now, we are ready to trace the evolution of the input state of $N$ independent particles injected into circuit in Fig.~\ref{Fig_Protocol_Wstate} which ends with post-selection in the output. Postponing the questions of experimental difficulties to the next section, like the phase stability which should be maintained during the evolution, we may write:
\begin{widetext}
\begin{eqnarray}\label{evolution-1}
\!\!\!\!a^{\dag}_{\scriptscriptstyle 1}\,a^{\dag}_{\scriptscriptstyle 2}\dots\,a^{\dag}_{\scriptscriptstyle N} \, \ket{0}
&\xymatrix{\ar[r]^{\atop U,G,V,...,V}_{Eqs.\,(\ref{U})\text{-}(\ref{G})\atop }  &}&\Big(\,\alpha\,a^{\dag}_{\scriptscriptstyle 1}+\tfrac{\beta}{\sqrt{N-1}}\,\big(\,a^{\dag}_{\scriptscriptstyle \dbloverline{1}}+.\,.\,.+a^{\dag}_{\scriptscriptstyle\dbloverline{N-1}}\,\big)\Big)\,\big(\,\delta\,a^{\dag}_{\scriptscriptstyle 2}+\varepsilon\,a^{\dag}_{\scriptscriptstyle \overline{2}}\,\big)\dots\big(\,\delta\,a^{\dag}_{\scriptscriptstyle N}+\varepsilon\,a^{\dag}_{\scriptscriptstyle \overline{N}}\,\big)\ket{0}
\\\label{evolution-2}
&\xymatrix{\ar[r]^{\atop \sigma}_{Eq.\,(\ref{sigma})\atop } &}&\Big(\,\alpha\,a^{\dag}_{\scriptscriptstyle 1}+\tfrac{\beta}{\sqrt{N-1}}\,\big(\,a^{\dag}_{\scriptscriptstyle 2}+.\,.\,.+a^{\dag}_{\scriptscriptstyle N}\,\big)\Big)\,\big(\,\delta\,a^{\dag}_{\scriptscriptstyle \dbloverline{1}}+\varepsilon\,a^{\dag}_{\scriptscriptstyle \overline{2}}\,\big)\dots\big(\,\delta\,a^{\dag}_{\scriptscriptstyle \dbloverline{N-1}}+\varepsilon\,a^{\dag}_{\scriptscriptstyle \overline{N}}\,\big)\ket{0}
\\\label{evolution-3}
&\xymatrix{\ar[r]^{\atop G^{\scriptscriptstyle\,\text{-}1}}_{Eq.\,(\ref{G-1})\atop } &}&\Big(\,\alpha\,a^{\dag}_{\scriptscriptstyle 1}+\tfrac{\beta}{\sqrt{N-1}}\,\big(\,a^{\dag}_{\scriptscriptstyle 2}+.\,.\,.+a^{\dag}_{\scriptscriptstyle N}\,\big)\Big)
\\\label{evolution-4}
&&\Big(\,\tfrac{\delta}{\sqrt{N-1}}\,\big(\,a^{\dag}_{\scriptscriptstyle \overline{1}}+\gamma^*_{\scriptscriptstyle 1,2}\,a^{\dag}_{\scriptscriptstyle \dbloverline{2}}+.\,.\,.+\gamma^*_{\scriptscriptstyle 1,N-1}\,a^{\dag}_{\scriptscriptstyle \dbloverline{N-1}}\,\big)+\varepsilon\,a^{\dag}_{\scriptscriptstyle \overline{2}}\,\Big)
\\\nonumber
&&\qquad\qquad\qquad\qquad\qquad\atop \cdot\cdot\cdot_{\ }
\\\label{evolution-5}
&&\Big(\,\tfrac{\delta}{\sqrt{N-1}}\,\big(\,a^{\dag}_{\scriptscriptstyle \overline{1}}+\gamma^*_{\scriptscriptstyle N-1,2}\,a^{\dag}_{\scriptscriptstyle \dbloverline{2}}+.\,.\,.+\gamma^*_{\scriptscriptstyle N-1,N-1}\,a^{\dag}_{\scriptscriptstyle \dbloverline{N-1}}\,\big)+\varepsilon\,a^{\dag}_{\scriptscriptstyle \overline{N}}\,\Big)\ket{0}
\\\label{W-post-selection}
&\xrsquigarrow{\text{\!\tiny{\emph{post-select}}\!}}&\Big(\,\alpha\varepsilon^{\scriptscriptstyle N-1}\ a^{\dag}_{\scriptscriptstyle {1}}a^{\dag}_{\scriptscriptstyle \overline{2}}\dots\, a^{\dag}_{\scriptscriptstyle \overline{N}}\ +\ \tfrac{\beta\delta\varepsilon^{N-2}}{N-1}\,\sum_{\scriptscriptstyle k\,=\,2}^{\scriptscriptstyle N}\ a^{\dag}_{\scriptscriptstyle {k}}\,a^{\dag}_{\scriptscriptstyle \overline{2}}\dots\,a^{\dag}_{\scriptscriptstyle \overline{k-1}}\ a^{\dag}_{\scriptscriptstyle \overline{1}}\,a^{\dag}_{\scriptscriptstyle \overline{k+1}}\,\dots\, a^{\dag}_{\scriptscriptstyle \overline{N}}\,\Big)\ket{0}
\\\label{W-state-out}
&&\stackrel{{\atop Eq.\,(\ref{all-equal})}}{=}\sqrt{\tfrac{\delta^2\,(1-\delta^2)^{N-1}}{\delta^2+(N-1)^2\,(1-\delta^2)}}\,\Big(\ket{\uparrow\downarrow\downarrow...\downarrow}+
\ket{\downarrow\uparrow\downarrow...\downarrow}+.\,.\,.+
\ket{\downarrow...\downarrow\uparrow\downarrow}+\ket{\downarrow...\downarrow\downarrow\uparrow}\!\Big)\,.\qquad\ \
\end{eqnarray}\vspace{0.2cm}
\end{widetext}
Note that the product structure in Eqs.~(\ref{evolution-3})-(\ref{evolution-5}) allows for quick identification of terms that remain after post-selection. Since we are interested only in cases with a single particle in each subsystem $B_{\scriptscriptstyle k}=\big\{\,k\,,\overline{k}\,\big\}$, then for each creator in the first bracket Eq.~(\ref{evolution-3}) there is only one choice of creators in the remaining brackets Eq.~(\ref{evolution-4})-(\ref{evolution-5}) which fulfils the post-selection condition: a single particle (creator) in each subsystem $B_{\scriptscriptstyle k}$ (notice that the modes with double bar can be dropped altogether).  What is left are only the terms in Eq.~(\ref{W-post-selection}). The last equality in Eq.~(\ref{W-state-out}) holds for  bosons  when all coefficients are equal
\begin{eqnarray}\label{all-equal}
\alpha\varepsilon^{\scriptscriptstyle N-1}&=&\tfrac{\beta\delta\varepsilon^{N-2}}{N-1}\,,
\end{eqnarray}
which holds for the choice $\alpha^2=\tfrac{\delta^2}{\delta^2+(N-1)^2(1-\delta^2)}$. Thus we obtain the  $W_{\scriptscriptstyle N}$ state Eq.~(\ref{W-state}) in the output. We remark that for fermions all terms except the first one in Eq.~(\ref{W-state-out}) get the '$-$' sign which can be easily corrected by phase shift $e^{i\pi}$ in the first path (either in the input or output).

\begin{figure}
\centering
\includegraphics[width=1\columnwidth]{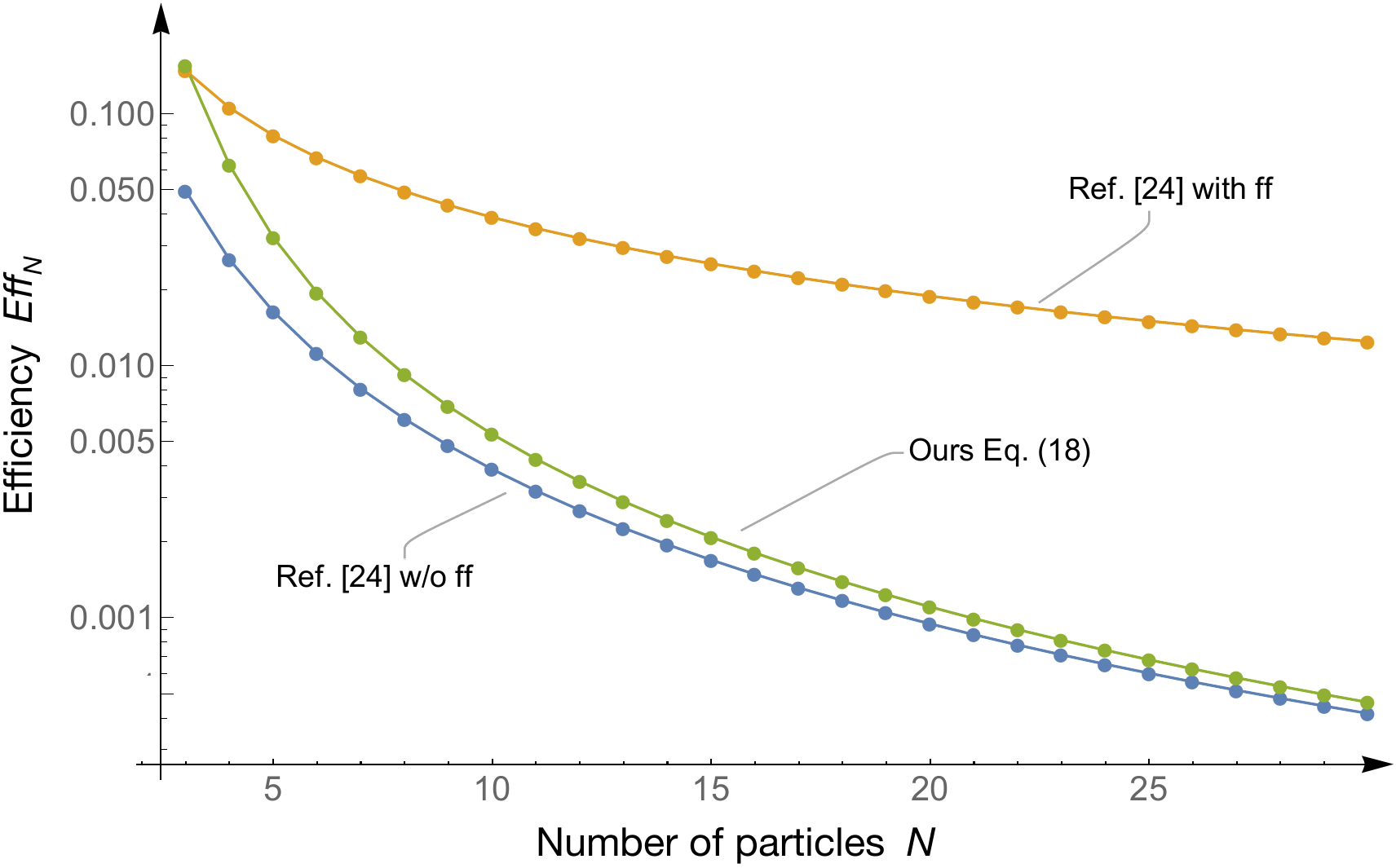}
\caption{\label{Fig_Efficiency_Wstate}{\bf\textsf{\mbox{Optimal efficiency of the scheme.}}} The success probability for generation of the $W_{\scriptscriptstyle N}$ state in our scheme Eq.~(\ref{Eff-N}) compared with the scheme with auxiliary particle and quantum erasure described in Ref.~\cite{KiChLiHa20} (without and with feedforward, see Eqs.~(12) and Eq.~(14) therein).}
\end{figure}

Note that the expression in Eq.~(\ref{W-state-out}) is unnormalised due to post-selection which projects on the subspace with a single particle in each channel $B_{\scriptscriptstyle 1}\,,...\,,B_{\scriptscriptstyle N}$. From the normalisation we can read out the success probability (efficiency) of the process which is equal to
\begin{eqnarray}\label{Eff-N-delta}
\textsl{\textsf{Eff}}_{\scriptscriptstyle N}(\delta)&=&\tfrac{N\,\delta^2\,(1-\delta^2)^{N-1}}{\delta^2+(N-1)^2\,(1-\delta^2)}\,.
\end{eqnarray}
Since there is one free parameter left in the protocol we can optimise over $\delta$ and obtain the maximal efficiency:
\begin{eqnarray}\label{Eff-N}
\textsl{\textsf{Eff}}_{\scriptscriptstyle N}&=&\max_{\scriptscriptstyle \delta}\,\textsl{\textsf{Eff}}_{\scriptscriptstyle N}(\delta)\,\sim\, \tfrac{e^{\scriptscriptstyle-1}}{N^{\scriptscriptstyle 2}}+\tfrac{7\,e^{\scriptscriptstyle-1}}{2\,N^{\scriptscriptstyle{3}}}+o\big(\tfrac{1}{N^{\scriptscriptstyle4}}\big)\,.
\end{eqnarray}
The result is plotted in Fig.~\ref{Fig_Efficiency_Wstate} and compared with the efficiency of the quantum erasure scheme in Ref.~\cite{KiChLiHa20}. See {\bf Appendix} for explicit calculation. Let us note that the obtained efficiency is insensitive to particle statistics, which is the same for bosons and fermions. This follows from two observations: \textit{(a)} in the passive linear optics regime the unitary evolution is governed by the same set of equations Eqs.~(\ref{U})-(\ref{G-1}) regardless of the statistics~\cite{Ti14}, and \textit{(b)} in the post-selected sector the same terms survive for both types of particles (due to the no-touching feature of the scheme~\cite{BlMa19}).

\begin{figure}
\centering
\includegraphics[width=\columnwidth]{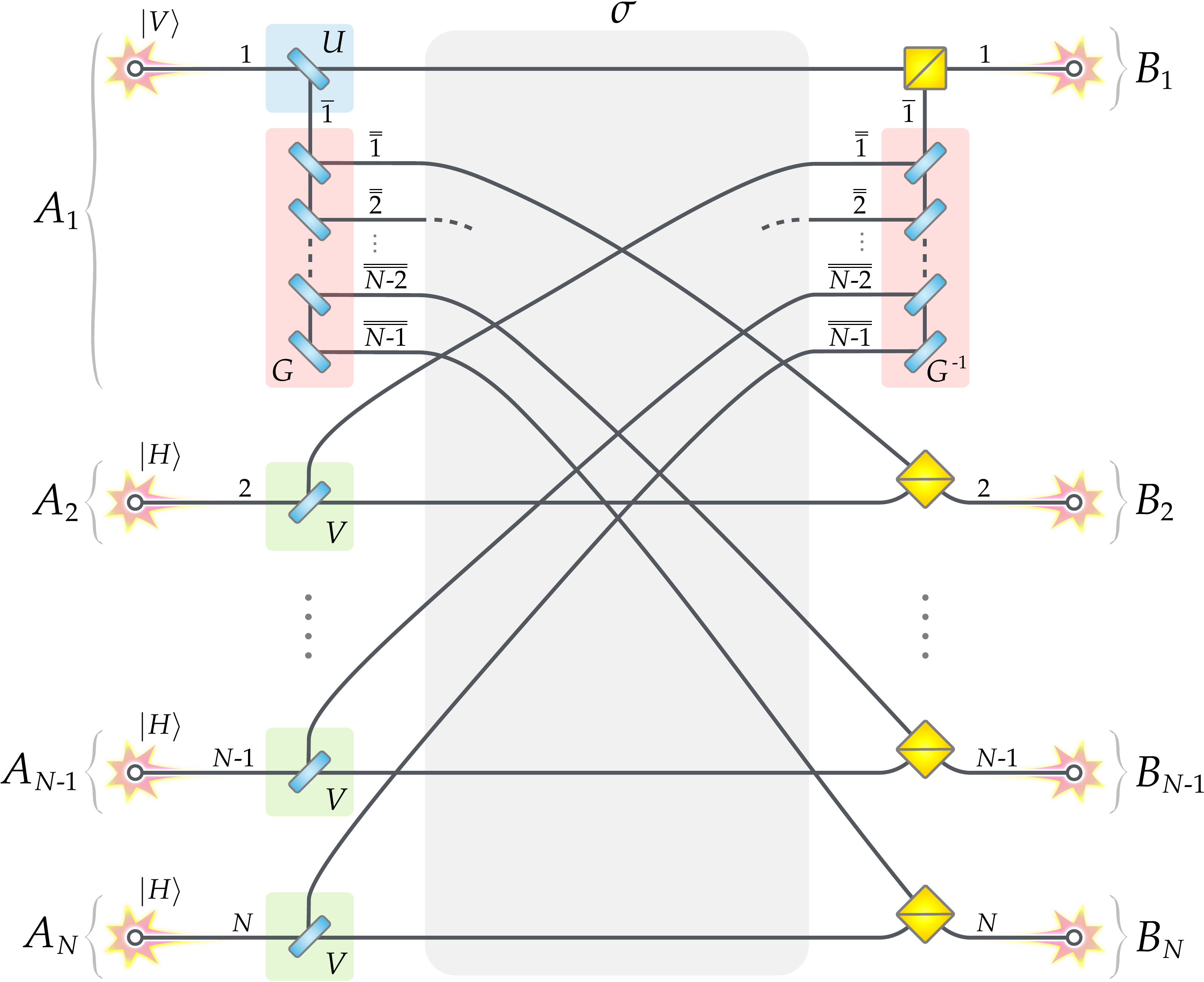}
\caption{\label{Fig_PolarisationEncoding_Wstate}{\bf\textsf{\mbox{Optical scheme with photon polarisation.}}} Rewriting of the proposal in Fig.~\ref{Fig_Protocol_Wstate} into photon polarisation qubits. In that case, each pair of paths $\{k\,,\overline{k}\}$ in dual-rail encoding gets replaced with two polarisation modes $\{\ket{H}_{\scriptscriptstyle k},\ket{V}_{\scriptscriptstyle k}\}$, with spatial merging obtained by polarising beam splitters (PBS) in the output. Then the unitaries $U$, $V$, $G$ and $G^{\scriptscriptstyle{-}1}$ are implemented by appropriate choice of beam splitters (BS). The analysis of this circuit parallels the dual-rail case in Fig.~\ref{Fig_Protocol_Wstate} and the resulting efficiencies are clearly the same.}
\end{figure}

We remark that the above discussion based on dual-rail encoding of qubits gives a generic pattern which directly translates into other realisations. For illustration, in Fig.~\ref{Fig_PolarisationEncoding_Wstate} we give analogous design for photon polarisation qubits using the scheme in Fig.~\ref{Fig_Protocol_Wstate} as a template.

\textit{Experimental feasibility.}---Our $N$ particle $W$ state generation protocol requires $N$ identical single photons and the phase stabilized linear optical networks. We remark that such technical demands can be fulfilled with the current quantum photonics technology. For example, it has been shown that tens of identical single photons can be generated either via spontaneous parametric down-conversion (SPDC)~\cite{ZhLiLiPeSuHuHe18} or from a quantum dot~\cite{WaQiDiChChYoHe19}. The temporal multiplexing technique with SPDC also provides a promising avenue to scale up the number of identical single photon generation~\cite{KaKw19}. The phase stabilized linear optical networks can be achieved with the active feedback control. For instance, a huge phase stabilized interferometer with an arm length of a few hundreds of kilometer has been implemented in the context of Twin-Field quantum key distribution~\cite{WaHeYiLuCuChZh19}. We also note that the integrated quantum photonics can provide an efficient way to implement complicated linear optical networks with the excellent phase stability~\cite{WaScLaTh20}. Therefore, our cost efficient $W$ state generation protocol can be implemented with tens of photons with the current quantum photonics technology.

\textit{Discussion.}---The presented protocol for generation of the $W$ state is cost-effective. It only requires linear optics and post-selection to get the $N$ particle $W$ state from $N$ independent particles in the input. The efficiency of state generation scales polynomially $\textsl{\textsf{Eff}}_{\scriptscriptstyle N}\sim\nicefrac{1}{N^2}$ with the number of particles. This leaves behind most proposals in the literature which use quantum fusion techniques~\cite{EiKiBoKuWe04,MiLiFuKo05,TaWaOzYaKoIm09,TaKiOzYaKoIm10,FaMeLiSiLo19} for the iterative construction of the larger $W_{\scriptscriptstyle N}$ states from smaller ones $W_{\scriptscriptstyle M<N}$, which inherently suffer form the exponential decrease of efficiency $\mathcal{O}^{\scriptscriptstyle-N}$. Our proposal clearly benefits from the direct construction of the state rather than use of iterative techniques.

Interestingly, there is a recent proposal of multipartite $W$ state generation based on quantum erasure~\cite{KiChLiHa20} (see also Ref.~\cite{BeLoCo17} for fermions). It is to our knowledge the most efficient protocol whose efficiency scales polynomially like $\nicefrac{1}{N^2}$, with a potential for further improvement to $\nicefrac{1}{N}$ by feed-forward with active state correction (see Fig~\ref{Fig_Efficiency_Wstate}). Note, however, that this scheme requires one auxiliary particle in the input on which appropriate measurement is made, i.e. $N+1$ particles are needed to produce the $N$ particle $W_{\scriptscriptstyle N}$ state. Clearly, this is an additional cost both in terms of particles and measurements that a fair comparison should take into account. Neither of those are required in our design and yet it performs slightly better (see {\bf Appendix}). As noted, even if post-processing based on feed-forward in the quantum erasure scheme in Ref.~\cite{KiChLiHa20} increases the efficiency, this comes with a considerable extra cost which scales linearly with the number of particles $N$ (and thus compromising the potential gain). 

Finally, we remark that the presented approach is insensitive to particle statistics, i.e. works with equal efficiency for fermions, bosons or anyons. This is a generic property of the no-touching designs~\cite{BlMa19}. Moreover, it can be safely used for Bell non-locality tests, since the 
post-selection scheme follows the all-but-one principle dsicussed in Ref.~\cite{BlBoMa20}, which assures no post-selection loophole. Therefore our protocol can be safely utilised as a part of device-independent protocols in quantum technological applications.

\textit{Acknowledgments.}---We thank M.~\.Zukowski for discussions and helpful comments. PB acknowledges support from the Polish National Agency for Academic Exchange in the Bekker Scholarship Programme. The work is part of the ICTQT IRAP (MAB) project of FNP, co-financed by structural funds of EU. YSK acknowledges support from the National Research Foundation of Korea (Grants No. 2019M3E4A1079777) and the KIST ORP program (2E31021).


\appendix

\section{\textsc{Appendix}}

Here we explicitly write out the formulas for the maximal efficiency $\textsl{\textsf{Eff}}_{\scriptscriptstyle N}$ in Eq.~(\ref{Eff-N}) for the scheme discussed in the paper.

Finding the maximum in Eq.~(\ref{Eff-N-delta}) is straightforward. It boils down to differentiating and choosing the solution such that $|\delta|\leqslant1$. Thus we get
\begin{eqnarray}
    \tfrac{\partial \textsl{\textsf{Eff}}_{\scriptscriptstyle N}(\delta)}{\partial \delta}\,=\,0&\Rightarrow&\delta_{\scriptscriptstyle max}^2\,=\,\tfrac{1-N+\sqrt{\frac{N^3-6N^2+13N-8}{N}}}{4-2N}\,,
\end{eqnarray} 
and hence the Eq.~(\ref{Eff-N}) takes the form 
\begin{eqnarray}
\textsl{\textsf{Eff}}_{\scriptscriptstyle N}\,=\,\textsl{\textsf{Eff}}_{\scriptscriptstyle N}(\delta_{\scriptscriptstyle max})\,=\,\tfrac{N\,\delta_{\scriptscriptstyle max}^2\,(1-\delta_{\scriptscriptstyle max}^2)^{N-1}}{\delta_{\scriptscriptstyle max}^2+(N-1)^2\,(1-\delta_{\scriptscriptstyle max}^2)}\,.
\end{eqnarray}
This result is plotted in Fig.~\ref{Fig_Efficiency_Wstate}.

For completeness, we also give asymptotic expansion of the above expression which reads
\begin{eqnarray}
\textsl{\textsf{Eff}}_{\scriptscriptstyle N}&\sim& \tfrac{e^{\scriptscriptstyle-1}}{N^{\scriptscriptstyle 2}}\,+\,\tfrac{7\,e^{\scriptscriptstyle-1}}{2\,N^{\scriptscriptstyle{3}}}\,+\,o\big(\tfrac{1}{N^{\scriptscriptstyle4}}\big)\,.
\end{eqnarray}
This should be compared with the corresponding efficiency for the recent proposal with quantum erasure in Ref.~\cite{KiChLiHa20} (to our knowledge the most efficient proposal for $W$ state generation in the literature). Eq.~(14) therein gives
\begin{eqnarray}
\textsl{\textsf{Eff}}_{\scriptscriptstyle N}^{\scriptscriptstyle \ (Kim\,et\,al)}&\sim& \tfrac{e^{\scriptscriptstyle-1}}{N^{\scriptscriptstyle 2}}\,+\,\tfrac{e^{\scriptscriptstyle-1}}{2\,N^{\scriptscriptstyle{3}}}\,+\,o\big(\tfrac{1}{N^{\scriptscriptstyle4}}\big)\,.
\end{eqnarray}
We note that although use of feedforward in the scheme of Kim et al~[5] increases the efficiency by factor $N$, it comes at an incomparable cost scaling with the number of particles $N$ too.

\bibliography{CombQuant}

\end{document}